\documentclass[12pt]{article}
\usepackage{graphicx,epsfig,psfig,color}
\newcommand{\bc}{\begin{center}}
\newcommand{\ec}{\end{center}}
\newcommand{\be}{\begin{equation}}
\newcommand{\ee}{\end{equation}}
\newcommand{\bea}{\begin{eqnarray}}
\newcommand{\eea}{\end{eqnarray}}
\newcommand{\ba}{\begin{array}}
\newcommand{\ea}{\end{array}}
\newcommand{\lb}{\label}
\newcommand{\rf}{\ref}
\newcommand{\bfg}{\begin{figure}[htbp]}
\newcommand{\efg}{\end{figure}}

\newcommand{\pr}{Phys. Rev. }
\newcommand{\np}{Nucl. Phys. }

\newcommand{\prp}{Phys. Rep. }

\newcommand{\pl}{Phys. Lett. }

\newcommand{\nc}{Nuovo Cimento }

\newcommand{\zp}{Z. Phys. }

\begin{document}

\begin{flushright}
IPNO-DR-05-06
\end{flushright}
\vspace{0.5 cm}
\bc
{\large \textbf{Quark-antiquark bound state equation in the \protect \\
Wilson loop approach with minimal surfaces\footnote{Talk given 
by H.S. at the Workshop QCD at Work 2005, Conversano, Italy,
16-20 June 2005.}}}
\vspace{1. cm}

F. Jugeau$^a$ and  H. Sazdjian$^b$\\
\vspace{0.25 cm}
\textit{$^a$Instituto de Fisica Corpuscular, IFIC,\\
Edificio Institutos de Investigacion, Apt. de Correus 22085,\\
E-46071 Val\`encia, Spain\\
\footnotesize{E-mail: frederic.jugeau@ific.uv.es}\\
\vspace{0.25 cm}
\normalsize
$^b$Institut de Physique Nucl\'eaire\footnote{Unit\'e Mixte de 
Recherche 8608.},
Groupe de Physique Th\'eorique,\\
Universit\'e Paris XI, F-91406 Orsay Cedex, France\\
\footnotesize{E-mail: sazdjian@ipno.in2p3.fr}}
\ec
\par
\renewcommand{\thefootnote}{\fnsymbol{footnote}}
\vspace{1.5 cm}

\bc
{\large Abstract}
\ec
\par
The quark-antiquark gauge invariant Green function is 
studied through its dependence on Wilson loops.
The latter are saturated, in the large-$N_c$ limit and
for large contours, by minimal surfaces.
A covariant bound state equation is derived which in the 
center-of-mass frame and at equal-times takes the form of
a Breit--Salpeter type equation. The large-distance 
interaction potentials reduce in the static case to a 
confining linear vector potential. In general, the interaction 
potentials involve contributions having the structure of 
flux tube like terms.
\par   
\vspace{0.5 cm}
\par
\newpage

The Wilson loop \cite{w} is defined as the trace in color space of the 
path-ordered phase factor of the gluon field on a closed contour $C$:
\be \lb{e2}
\Phi(C)=\frac{1}{N_c}\mathrm{tr}_cPe^{{\displaystyle 
-ig\oint_Cdx^{\mu}A_{\mu}(x)}}.
\ee
Its vacuum expectation value, denoted $W(C)$,
\be \lb{e3}
W(C)=\langle\Phi(C)\rangle,
\ee
is a functional of the contour $C$.
Loop equations were obtained and studied by Polyakov \cite{p} and Makeenko 
and Migdal \cite{mm,mgd,mk}. The Wilson loop essentially satisfies 
two types of equation, which are equivalent to the QCD
equations of motion: The Bianchi identity and the loop equations 
(or Makeenko--Migdal equations). A third property, factorization, 
is obtained in the large-$N_c$ limit \cite{th} for two disjoint contours:
$W(C_1,C_2)=W(C_1)W(C_2)$.
\par
Further simplification is obtained in the large-$N_c$ limit of
the theory. In that limit, for large contours, i.e., at large
distances, nonperturbative asymptotic solutions to the Wilson
loops are represented by the minimal surfaces having as supports
the loop contours \cite{mm,js}. Therefore, if one is interested only 
in the large-distance properties of the theory, saturation of the Wilson 
loop averages by minimal surfaces provides a correct description of the 
theory in this regime. In that case, the Wilson loop average can be 
represented by the following functional of the contour $C$:
\be \lb{e5}
W(C)=e^{{\displaystyle -i\sigma A(C)}},
\ee
where $\sigma$ is the string tension and $A(C)$ the minimal area
with contour $C$.
\par
Minimal surfaces also appear as natural solutions to the Wilson
loop averages in two-dimensional gauge theories \cite{kkk}. 
\par
To deal with the quarkonium bound state problem, one starts 
with the two-particle gauge invariant Green function for
quarks $q_1$ and $q_2$ with different flavors and with masses 
$m_1$ and $m_2$:
\be \lb{e6}
G(x_1,x_2;x_1',x_2')\equiv \langle \overline \psi_2(x_2)U(x_2,x_1)
\psi_1(x_1)\overline \psi_1(x_1')U(x_1',x_2')\psi_2(x_2')
\rangle_{A,q_1,q_2}.
\ee
Here, $U(x_2,x_1)$ is the path-ordered phase factor, 
\be \lb{e1}
U(x_2,x_1)=Pe^{{\displaystyle -ig\int_{x_1}^{x_2} 
dz^{\mu}A_{\mu}(z)}},
\ee
taken along the straight-line $x_1x_2$ (and similarly for $U(x_1',x_2')$).
Integrating in the large-$N_c$ limit with respect to the
quark fields, one obtains:
\be \lb{e7}
G(x_1,x_2;x_1',x_2')=-\langle \mathrm{tr}_c\,U(x_2,x_1)S_1(A;x_1,x_1')
U(x_1',x_2')S_2(A;x_2',x_2)\rangle_A,
\ee
where $S_1(A)$ and $S_2(A)$ are the quark and antiquark propagators in the 
presence of the external gluon field and $\mathrm{tr}_c$ designates the 
trace with respect to the color group. The quark propagator $S(A)$
satisfies the equation
\be \lb{e7a}
\Big(i\gamma.\partial_x-m-g\gamma.A(x)\Big)S(A;x,x')=i\delta^4(x-x').
\ee
\par
At this stage, the Green function $G$ can schematically be represented
as in Fig. \rf{f1}. 
\bfg
\bc
\vspace*{0.5 cm}
\begin{picture}(0,0)%
\epsfig{file=f1.pstex}%
\end{picture}%
\setlength{\unitlength}{2368sp}%
\begingroup\makeatletter\ifx\SetFigFont\undefined%
\gdef\SetFigFont#1#2#3#4#5{%
  \reset@font\fontsize{#1}{#2pt}%
  \fontfamily{#3}\fontseries{#4}\fontshape{#5}%
  \selectfont}%
\fi\endgroup%
\begin{picture}(4812,2568)(2551,-4399)
\put(3151,-2011){\makebox(0,0)[lb]{\smash{\SetFigFont{9}{10.8}{\familydefault}{\mddefault}{\updefault}{\color[rgb]{0,0,0}$x_1$}%
}}}
\put(6826,-2011){\makebox(0,0)[lb]{\smash{\SetFigFont{9}{10.8}{\familydefault}{\mddefault}{\updefault}{\color[rgb]{0,0,0}$x_1'$}%
}}}
\put(2551,-4336){\makebox(0,0)[lb]{\smash{\SetFigFont{9}{10.8}{\familydefault}{\mddefault}{\updefault}{\color[rgb]{0,0,0}$x_2$}%
}}}
\put(7126,-4111){\makebox(0,0)[lb]{\smash{\SetFigFont{9}{10.8}{\familydefault}{\mddefault}{\updefault}{\color[rgb]{0,0,0}$x_2'$}%
}}}
\put(7126,-2911){\makebox(0,0)[lb]{\smash{\SetFigFont{9}{10.8}{\familydefault}{\mddefault}{\updefault}{\color[rgb]{0,0,0}$U$}%
}}}
\put(4801,-4186){\makebox(0,0)[lb]{\smash{\SetFigFont{9}{10.8}{\familydefault}{\mddefault}{\updefault}{\color[rgb]{0,0,0}$S_2(A)$}%
}}}
\put(4876,-2011){\makebox(0,0)[lb]{\smash{\SetFigFont{9}{10.8}{\familydefault}{\mddefault}{\updefault}{\color[rgb]{0,0,0}$S_1(A)$}%
}}}
\put(2776,-2986){\makebox(0,0)[lb]{\smash{\SetFigFont{9}{10.8}{\familydefault}{\mddefault}{\updefault}{\color[rgb]{0,0,0}$U$}%
}}}
\end{picture}

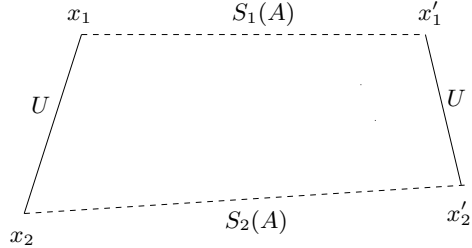
\caption{Schematic representation of the two-particle Green
function.}
\lb{f1}
\ec
\efg
\par
In order to make the Wilson loop structure of $G$ 
apparent, we adopt for the quark propagator in the external gluon field 
a representation based on an explicit use of the phase
factor along straight lines \cite{js}. Introducing the gauge
covariant composite object $\widetilde S(A;x,x')$, made of a free fermion 
propagator $S_0(x-x')$ (without color group content) multiplied by the 
path-ordered phase factor $U(x,x')$ [Eq. (\rf{e1})] taken along the 
straight segment $x'x$,
\be \lb{e9}
\widetilde S(A;x,x')\equiv S_0(x-x')U(x,x'),
\ee
one shows that the quark propagator $S(A;x,x')$ in the external
gluon field satisfies the following functional integral equation in 
terms of $\widetilde S$:
\be \lb{e10}
S(A;x,x')=\widetilde S(A;x,x')-\int d^4x''S(A;x,x'')\gamma^{\alpha}
\int_0^1d\lambda\,(1-\lambda)\frac{\delta}{\delta x^{\alpha}(\lambda)}
\widetilde S(A;x'',x'),
\ee
where the segment $x''x'$ has been parametrized with the parameter
$\lambda$ as $x(\lambda)=(1-\lambda)x''+\lambda x'$ and 
where the operator $\delta/\delta x^{\alpha}(\lambda)$ acts on the
factor $U$ of $\widetilde S$, along the internal part of the segment 
$x''x'$, with $x'$ held fixed. That equation is diagrammatically 
represented in Fig. \rf{f2}.
\par
\bfg
\bc
\vspace*{0.5 cm}
\begin{picture}(0,0)%
\epsfig{file=f2.pstex}%
\end{picture}%
\setlength{\unitlength}{2368sp}%
\begingroup\makeatletter\ifx\SetFigFont\undefined%
\gdef\SetFigFont#1#2#3#4#5{%
  \reset@font\fontsize{#1}{#2pt}%
  \fontfamily{#3}\fontseries{#4}\fontshape{#5}%
  \selectfont}%
\fi\endgroup%
\begin{picture}(8166,1077)(2657,-4094)
\put(4801,-4036){\makebox(0,0)[lb]{\smash{\SetFigFont{9}{10.8}{\familydefault}{\mddefault}{\updefault}{\color[rgb]{0,0,0}$=$}%
}}}
\put(7426,-4036){\makebox(0,0)[lb]{\smash{\SetFigFont{9}{10.8}{\familydefault}{\mddefault}{\updefault}{\color[rgb]{0,0,0}$+$}%
}}}
\put(6151,-3811){\makebox(0,0)[lb]{\smash{\SetFigFont{9}{10.8}{\familydefault}{\mddefault}{\updefault}{\color[rgb]{0,0,0}$\widetilde S(A)$}%
}}}
\put(10501,-3661){\makebox(0,0)[lb]{\smash{\SetFigFont{9}{10.8}{\familydefault}{\mddefault}{\updefault}{\color[rgb]{0,0,0}$\widetilde S(A)$}%
}}}
\put(3451,-3811){\makebox(0,0)[lb]{\smash{\SetFigFont{9}{10.8}{\familydefault}{\mddefault}{\updefault}{\color[rgb]{0,0,0}$S(A)$}%
}}}
\put(8176,-3361){\makebox(0,0)[lb]{\smash{\SetFigFont{9}{10.8}{\familydefault}{\mddefault}{\updefault}{\color[rgb]{0,0,0}$S(A)$}%
}}}
\end{picture}

\caption{Diagrammatic representation of the integral equation
satisfied by the quark propagator in the external gluon 
field. The cross represents the action of the functional derivative
$\delta/\delta x(\lambda)$.}
\lb{f2}
\ec
\efg
A similar equation in which the roles of $x$ and $x'$ are interchanged 
also holds. Those equations lead to iteration series for $S$ in which 
the gauge covariance property is maintained at each order of the iteration. 
\par
Use of the above representations for the quark propagators in Eq.
(\rf{e7}) leads for the two-particle Green function to a series
expansion where each term contains a Wilson loop along a skew-polygon:
\be \lb{e11}
G=\sum_{i,j=1}^{\infty}G_{i,j},
\ee
where $G_{i,j}$ represents the contribution of the term of the series
having $(i-1)$ points of integration between $x_1$ and $x_1'$ ($i$
segments) and $(j-1)$ points of integration between $x_2$ and $x_2'$
($j$ segments). We designate by $C_{i,j}$ the contour associated
with the term $G_{i,j}$. A typical configuration for the contour of
$G_{4,3}$ is represented in Fig. \rf{f3}.
\par
\bfg
\bc
\vspace*{0.5 cm}
\begin{picture}(0,0)%
\includegraphics{f3.pstex}%
\end{picture}%
\setlength{\unitlength}{2960sp}%
\begingroup\makeatletter\ifx\SetFigFont\undefined%
\gdef\SetFigFont#1#2#3#4#5{%
  \reset@font\fontsize{#1}{#2pt}%
  \fontfamily{#3}\fontseries{#4}\fontshape{#5}%
  \selectfont}%
\fi\endgroup%
\begin{picture}(6249,3105)(2464,-4174)
\put(5626,-1261){\makebox(0,0)[lb]{\smash{\SetFigFont{11}{13.2}{\familydefault}{\mddefault}{\updefault}{\color[rgb]{0,0,0}$y_2$}%
}}}
\put(3151,-2086){\makebox(0,0)[lb]{\smash{\SetFigFont{11}{13.2}{\familydefault}{\mddefault}{\updefault}{\color[rgb]{0,0,0}$x_1$}%
}}}
\put(6301,-2011){\makebox(0,0)[lb]{\smash{\SetFigFont{11}{13.2}{\familydefault}{\mddefault}{\updefault}{\color[rgb]{0,0,0}$y_3$}%
}}}
\put(6376,-4111){\makebox(0,0)[lb]{\smash{\SetFigFont{11}{13.2}{\familydefault}{\mddefault}{\updefault}{\color[rgb]{0,0,0}$z_2$}%
}}}
\put(4801,-2461){\makebox(0,0)[lb]{\smash{\SetFigFont{11}{13.2}{\familydefault}{\mddefault}{\updefault}{\color[rgb]{0,0,0}$A_{4,3}$}%
}}}
\put(3151,-3661){\makebox(0,0)[lb]{\smash{\SetFigFont{11}{13.2}{\familydefault}{\mddefault}{\updefault}{\color[rgb]{0,0,0}$x_2$}%
}}}
\put(4726,-3361){\makebox(0,0)[lb]{\smash{\SetFigFont{11}{13.2}{\familydefault}{\mddefault}{\updefault}{\color[rgb]{0,0,0}$z_1$}%
}}}
\put(7276,-3436){\makebox(0,0)[lb]{\smash{\SetFigFont{11}{13.2}{\familydefault}{\mddefault}{\updefault}{\color[rgb]{0,0,0}$x_2'$}%
}}}
\put(7501,-2011){\makebox(0,0)[lb]{\smash{\SetFigFont{11}{13.2}{\familydefault}{\mddefault}{\updefault}{\color[rgb]{0,0,0}$x_1'$}%
}}}
\put(4426,-1411){\makebox(0,0)[lb]{\smash{\SetFigFont{11}{13.2}{\familydefault}{\mddefault}{\updefault}{\color[rgb]{0,0,0}$y_1$}%
}}}
\end{picture}

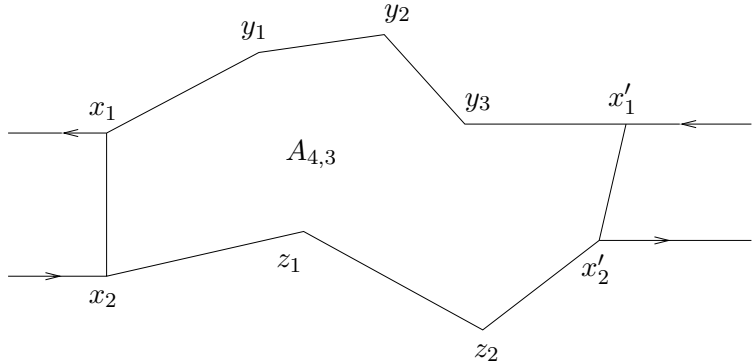
\caption{Contour $C_{4.3}$ associated with the term $G_{4,3}$.
$A_{4,3}$ is the minimal surface with contour $C_{4,3}$.}
\lb{f3}
\ec
\efg
Each segment of the quark lines supports a free quark propagator and 
except for the first segments (or the last ones, depending on the 
representation that is used) the Wilson loop is submitted to one 
functional derivative on each such segment. One then uses for the 
averages of the Wilson loops appearing in the above series the 
representation with minimal surfaces [Eq. (\rf{e5})].
\par
The Green function $G$ satisfies the following equation with respect to
the Dirac operator of particle 1 acting on $x_1$:
\bea \lb{e8}
& &(i\gamma.\partial_{(x_1)}-m_1)G(x_1,x_2;x_1',x_2')=
-i\langle\mathrm{tr}_c\,U(x_2,x_1)\delta^4(x_1-x_1')U(x_1',x_2')
S_2(x_2',x_2)\rangle_A\nonumber \\
& &\ \ \ \ \ \ -i\gamma^{\alpha}\langle\mathrm{tr}_c\int_0^1 
d\sigma(1-\sigma)\frac{\delta U(x_2,x_1)}{\delta x^{\alpha}(\sigma)}
S_1(x_1,x_1')U(x_1',x_2')S_2(x_2',x_2)\rangle_A,
\eea
where the segment $x_1x_2$ has been parametrized with the parameter
$\sigma$ as $x(\sigma)=(1-\sigma)x_1+\sigma x_2$; furthermore, the
operator $\delta/\delta x^{\alpha}$ does not act on the explicit
boundary point $x_1$  of the segment, this contribution having
been cancelled by the contribution of the gluon field $A$ coming from
the quark propagator $S_1$. A similar equation also holds with the Dirac
operator of particle 2 acting on $x_2$.
Representation (\rf{e10}) for the quark propagator can then be used in
the above equation and its partner satisfied by the two-particle
Green function $G$. One obtains two compatible equations for
$G$ where the right-hand sides involve the series of the terms
$G_{i,j}$ of Eq. (\rf{e11}) and their functional derivative along
the segment $x_1x_2$. In order to obtain bound state equations,
it is necessary to reconstruct in the right-hand sides the bound
state poles contained in $G$ \cite{sb}. In $x$-space, bound states
are reached by taking the large separation time limit between the pair
of points ($x_1,x_2$) and ($x_1',x_2'$) \cite{gml}.
To produce a bound state pole, it is necessary that there be a coherent
sum of contributions coming from each $G_{i,j}$, since the latter,
taken individually, do not have poles.
Each $G_{i,j}$ involves a corresponding minimal surface $A_{i,j}$ on
the contour of which act various functional derivatives. Those can be
classified according to their possible irreducibility properties.
Reducible contributions are those which are parts of the definition
of the series of $G$. It does not seem possible to sum all these
terms to reproduce exactly $G$ with some kernel acting on it in the
right-hand sides. However, for large separation time limits one
can isolate terms that contribute to the pole terms.
One notices that the derivative along the segment $x_1x_2$ acts on areas 
$A_{i,j}$ with contour $C_{i,j}$ which are different from one term 
of the series to the other (the number of segments being different). 
To have a coherent sum of those contributions it is necessary to 
expand each such derivative term around the derivative of the lowest-order 
contour $C_{1,1}$, represented in Fig. \rf{f4}.
\par
\bfg
\bc
\vspace*{0.5 cm}
\begin{picture}(0,0)%
\epsfig{file=f4.pstex}%
\end{picture}%
\setlength{\unitlength}{2960sp}%
\begingroup\makeatletter\ifx\SetFigFont\undefined%
\gdef\SetFigFont#1#2#3#4#5{%
  \reset@font\fontsize{#1}{#2pt}%
  \fontfamily{#3}\fontseries{#4}\fontshape{#5}%
  \selectfont}%
\fi\endgroup%
\begin{picture}(4662,2406)(2326,-3949)
\put(4201,-2761){\makebox(0,0)[lb]{\smash{\SetFigFont{11}{13.2}{\familydefault}{\mddefault}{\updefault}{\color[rgb]{0,0,0}$A_{1,1}$}%
}}}
\put(2326,-3661){\makebox(0,0)[lb]{\smash{\SetFigFont{11}{13.2}{\familydefault}{\mddefault}{\updefault}{\color[rgb]{0,0,0}$x_2$}%
}}}
\put(6901,-3886){\makebox(0,0)[lb]{\smash{\SetFigFont{11}{13.2}{\familydefault}{\mddefault}{\updefault}{\color[rgb]{0,0,0}$x_2'$}%
}}}
\put(6451,-1711){\makebox(0,0)[lb]{\smash{\SetFigFont{11}{13.2}{\familydefault}{\mddefault}{\updefault}{\color[rgb]{0,0,0}$x_1'$}%
}}}
\put(2476,-2011){\makebox(0,0)[lb]{\smash{\SetFigFont{11}{13.2}{\familydefault}{\mddefault}{\updefault}{\color[rgb]{0,0,0}$x_1$}%
}}}
\end{picture}

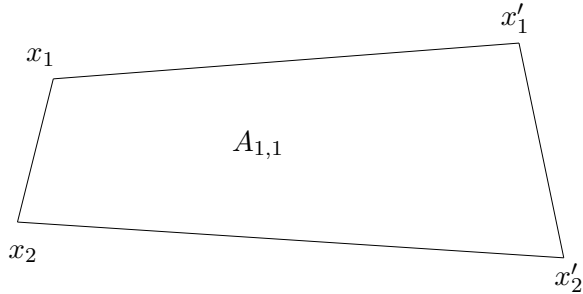
\caption{The lowest-order contour $C_{1,1}$ and its minimal surface
$A_{1,1}$.}
\lb{f4}
\ec
\efg
It is that term that can be factorized and can lead through the
summation of the factored series to the reappearance of the
Green function $G$ and to its poles. The remaining terms do not
lead to pole terms. Similarly, two derivative contributions  
should be expanded around the lowest-order contribution
coming from the contours $C_{2,1}$ or $C_{1,2}$, and so forth.
\par
In general, the derivative of the areas along $x_1x_2$ depends
among others on the slope of the areas in the orthogonal direction 
to $x_1x_2$. One then associates that slope with the quark momenta.
Taking then the large separation time limit and equal times in the
center-of-mass frame, one ends up with a covariant 
three-dimensional equation, having the structure of a
Breit--Salpeter type equation \cite{b,s} and where the 
interaction kernels or potentials are given by various functional
derivatives involving at least one derivative along the segment 
$x_1x_2$. Keeping for the potentials the terms containing one functional 
derivative of the area $A_{1,1}$ [Fig. \rf{f1}], the equation takes the 
form \cite{js} 
\be \lb{e12}
\Big[P_0-(h_{10}+h_{20})-\gamma_{10}\gamma_1^{\mu}A_{1\mu}
-\gamma_{20}\gamma_2^{\mu}A_{2\mu}\Big]\psi(\mathbf{x})=0,
\ee
where $\psi$ is a $4\times 4$ matrix wave function of the relative 
coordinate $x=x_2-x_1$ considered at equal times, $P_0$ the 
center-of-mass total energy and $h_{10}$ and $h_{20}$ the quark and 
antiquark Dirac hamiltonians; the Dirac matrices of the quark (with 
index 1) act on $\psi$ from the left, while the Dirac matrices of the 
antiquark (with index 2) act on $\psi$ from the right. The potentials 
$A_1$ and $A_2$ are defined through the equations 
\be \lb{e13}
A_{1\mu}=\sigma\int_0^1d\sigma'(1-\sigma')\frac{\delta A_{1,1}}
{\delta x^{\mu}(\sigma')},\ \ \ \ \ 
A_{2\mu}=\sigma\int_0^1d\sigma'\,\sigma'\frac{\delta A_{1,1}}
{\delta x^{\mu}(\sigma')},
\ee
$x(\sigma')$ belonging to the segment $x_1x_2$.
\par
The time components of $A_1$ and $A_2$ add up in the wave
equation. For their sum, one has the expression (in the c.m. frame)
\bea \lb{e14}
& &A_{10}+A_{20}=\sigma r\frac{E_1E_2}{E_1+E_2}
\bigg\{\Big(\frac{E_1}{E_1+E_2}\epsilon(p_{10})+
\frac{E_2}{E_1+E_2}\epsilon(p_{20})\Big)\nonumber \\
& &\ \ \ \ \ \ \ \ \times\sqrt{\frac{r^2}{\mathbf{L}^2}}
\left(\,\arcsin\Big(\frac{1}{E_2}\sqrt{\frac{\mathbf{L}^2}{r^2}}\Big)+
\arcsin\Big(\frac{1}{E_1}\sqrt{\frac{\mathbf{L}^2}{r^2}}\Big)\,\right)
\nonumber \\
& &+(\epsilon(p_{10})-\epsilon(p_{20}))
\Big(\frac{E_1E_2}{E_1+E_2}\Big)\Big(\frac{r^2}{\mathbf{L}^2}\Big)
\left(\,\sqrt{1-\frac{\mathbf{L}^2}{r^2E_2^2}}-
\sqrt{1-\frac{\mathbf{L}^2}{r^2E_1^2}}\,\right)\bigg\}.\nonumber \\
& &
\eea
Here, $r=\sqrt{\mathbf{x}^2}$,
$E_a=\sqrt{m_a^2+\mathbf{p}^2}$, $a=1,2$, with $m_a$ the quark
masses, $\mathbf{p}$ the c.m. momentum, 
$\mathbf{p}=(\mathbf{p}_2-\mathbf{p}_1)/2$, $\mathbf{L}$ the c.m.
orbital angular momentum, and $\epsilon(p_{10})$ and $\epsilon(p_{20})$
the energy sign operators of the free quark and the antiquark, 
respectively:
\be \lb{e15}
\epsilon(p_{a0})=\frac{h_{a0}}{E_a},\ \ \ \ \ a=1,2.
\ee
\par 
The space components of $A_1$ and $A_2$ are orthogonal to 
$\mathbf{x}$. The expression of $\mathbf{A}_1$ is (in the c.m. frame):
\bea \lb{e16}
\mathbf{A}_1&=&-\sigma r\frac{E_1E_2}{E_1+E_2}
\bigg\{\,\frac{r^2}{2\mathbf{L}^2}\frac{E_1E_2}{E_1+E_2}\mathbf{p}^t
\nonumber \\
& &\ \ \ \ \ \times\sqrt{\frac{r^2}{\mathbf{L}^2}}
\left(\,\arcsin\Big(\frac{1}{E_2}\sqrt{\frac{\mathbf{L}^2}{r^2}}\Big)+
\arcsin\Big(\frac{1}{E_1}\sqrt{\frac{\mathbf{L}^2}{r^2}}\Big)\,\right)
\nonumber \\
& &\ \ \ +\frac{1}{E_2}\mathbf{p}^t\Big(\frac{E_1E_2}{E_1+E_2}\Big)
\Big(\frac{r^2}{\mathbf{L}^2}\Big)
\left(\,\sqrt{1-\frac{\mathbf{L}^2}{r^2E_2^2}}-
\sqrt{1-\frac{\mathbf{L}^2}{r^2E_1^2}}\,\right)\nonumber \\
& &\ \ \ -\frac{1}{2}\mathbf{p}^t\Big(\frac{r^2}{\mathbf{L}^2}\Big)
\left(\,\frac{E_1}{E_1+E_2}\sqrt{1-
\frac{\mathbf{L}^2}{r^2E_2^2}}+\frac{E_2}{E_1+E_2}
\sqrt{1-\frac{\mathbf{L}^2}{r^2E_1^2}}\,\right)\,\bigg\}.
\eea
Here, $\mathbf{p}^t$ is the transverse part of $\mathbf{p}$ with
respect to $\mathbf{x}$:
\be \lb{e17}
\mathbf{p}^t=\mathbf{p}-\mathbf{x}\frac{1}{\mathbf{x}^2}
\mathbf{x}.\mathbf{p}.
\ee
The expression of $\mathbf{A}_2$ is obtained from that of 
$\mathbf{A}_1$ by an interchange in the latter of the indices 1 and 2 
and a change of sign of $\mathbf{p}^t$.  
\par
For sectors of quantum numbers where $\mathbf{L}^2=0$, the expressions 
of the potentials become:
\bea 
\lb{e18}
& &A_{10}+A_{20}=\frac{1}{2}(\epsilon(p_{10})+\epsilon(p_{20}))
\sigma r,\\
\lb{e19}
& &\mathbf{A}_1=-\frac{1}{E_1E_2}\Big(\frac{1}{3}(E_1+E_2)-
\frac{1}{2}E_1\Big)\mathbf{p}^t\sigma r,\nonumber \\
& &\mathbf{A}_2=+\frac{1}{E_1E_2}\Big(\frac{1}{3}(E_1+E_2)-
\frac{1}{2}E_2\Big)\mathbf{p}^t\sigma r.
\eea
\par
The potentials are generally momentum dependent operators and 
necessitate an appropriate ordering of terms.
\par
From the structure of the wave equation (\rf{e12}) and the expressions 
of the potentials, one deduces that the interaction is confining and 
of the vector type. However, compared to the conventional timelike
vector potential, it has additional pieces of terms contributing
to the orbital angular momentum dependent parts. A closer analysis
of those terms shows that they can be interpreted as being originated
from the moments of inertia of the segment $x_1x_2$ carrying a constant
linear energy density equal to the string tension. The interaction
potentials are therefore provided by the energy-momentum vector of the
segment joining the quark to the antiquark, in similarity with the
color flux tube picture of confinement. An analogous equation had
also been proposed by Olsson \textit{et al.} on the basis of a model
where the quarks are attached at the ends of a straight string or a
color flux tube \cite{ow,lcooowoo}. A similar conclusion had also been
reached by Brambilla, Prosperi \textit{et al.} on the basis of the
analysis of the relativistic corrections to the nonrelativistic limit
of the Wilson loop \cite{bmpbbp,bcpbmp,bp}.
\par
For heavy quarks, one can expand equation (\rf{e12}) around the
nonrelativistic limit and obtain the hamiltonian to order $1/c^2$
\cite{js}.
\par
The relativistic corrections to the interquark potential arising
from the Wilson loop were analyzed and evaluated in the literature by 
Eichten and Feinberg \cite{ef}, Gromes \cite{gr}, Brambilla, Prosperi
\textit{et al.} \cite{bmpbbp,bcpbmp,bp}, Brambilla, Pineda, Soto and 
Vairo \cite{pvbpsv}.
\par
The Wilson loop approach was also used for the study of quarkonium 
systems by Dosch, Simonov \textit{et al.} with the use of the
stochastic vacuum model \cite{dssdg}.
\par
In conclusion, the saturation of the Wilson loop averages in the 
large-$N_c$ limit by minimal surfaces provides a systematic tool for 
investigating the large-distance dynamics of quark-antiquark bound
state systems.
\par

\vspace{0.25 cm}

\vspace{0.25 cm}
\noindent
\textbf{Acknowledgements}: This work was supported in part for H.S. 
by the EU RTN network EURIDICE under contract No. CT2002-0311.
\par

\end{document}